# Hemodynamics of Stent Implantation Procedures in Coronary Bifurcations: an in vitro study

*Melissa C. Brindise[1], Claudio Chiastra[2,3], Francesco Burzotta[4],*

*Francesco Migliavacca[2], and Pavlos P. Vlachos[1]*

[1] School of Mechanical Engineering, Purdue University, West Lafayette, IN, USA

[2] Laboratory of Biological Structure Mechanics (LaBS), Chemistry, Materials and Chemical Engineering Department "Giulio Natta", Politecnico di Milano, Milan, Italy

[3] Department of Biomedical Engineering, Thoraxcenter, Erasmus University Medical Center, Rotterdam, The Netherlands

[4] Department of Cardiovascular Medicine, Università Cattolica del Sacro Cuore, Rome, Italy

**Address correspondence:**
Pavlos P. Vlachos,
School of Mechanical Engineering, Purdue University, 585 Purdue Mall, West Lafayette, IN, 47907.
Electronic mail: pvlachos@purdue.edu

# Abstract

Stent implantation in coronary bifurcations presents unique challenges and currently there is no universally accepted stent deployment approach. Despite clinical and computational studies, to date, the effect of each stent implantation method on the coronary artery hemodynamics is not well understood. In this study the hemodynamics of stented coronary bifurcations under pulsatile flow conditions were investigated experimentally. Three implantation methods, provisional side branch (PSB), culotte (CUL), and crush (CRU), were investigated using time-resolved particle image velocimetry (PIV) to measure the velocity fields. Subsequently, hemodynamic parameters including wall shear stress (WSS), oscillatory shear index (OSI), and relative residence time (RRT) were calculated and the pressure field through the vessel was non-invasively quantified. The effects of each stented case were evaluated and compared against an un-stented case. CRU provided the lowest compliance mismatch, but demonstrated detrimental stent interactions. PSB, the clinically preferred method, and CUL maintained many normal flow conditions. However, PSB provided about a 300% increase in both OSI and RRT. CUL yielded a 10% and 85% increase in OSI and RRT, respectively. The results of this study support the concept that different bifurcation stenting techniques result in hemodynamic environments that deviate from that of un-stented bifurcations, to varying degrees.

# 1. Introduction

Percutaneous coronary interventions on bifurcated coronary lesions represent a challenge for interventional cardiologists. Though a series of different stent implantation techniques have been described, uncertainty still exists regarding the best selection for each individual patient.[21]

The (drug-eluting) stent implantation procedure aims to minimize the occurrence of both vessel thrombosis and in-stent restenosis which are the main phenomena causing major adverse cardiac events (MACE). The main stent failures (restenosis and thrombosis) have been associated with the flow dynamics of stented segments, thus calling for improvements in the assessment and minimization of local stent-induced hemodynamic changes.[15,1] Flow parameters that have a proven effect on stent implantation include wall shear stress (WSS), oscillatory shear index (OSI), and relative residence time (RRT). Decreases in WSS values, as well as increases in OSI and RRT increase the risk of restenosis.[20,17,1] Compliance mismatch between the stent and the host vessel can also increase the risk of thrombosis.[28]

Implantation techniques for coronary bifurcations can utilize one or two stents.[21] The simplest stent technique is the provisional side branch (PSB) method, which uses only one stent in the main vessel (MV). It is often eventually followed by further interventions (like ballooning or stenting) in the side branch (SB) through the stent struts. Conversely, double stenting strategies deploy stents in both the MV and the SB using various techniques. Among different double stenting techniques, the culotte (CUL) and the crush (CRU) have been widely adopted worldwide.

Clinical trials have been a primary method for evaluating stent implantation techniques. PSB is currently the preferred method largely due to its simplicity, and easier and shorter implantation procedure.[8] Short-term clinical trials have suggested that the PSB method produces less adverse events (8.0%) as compared to complex stenting techniques (15.2%)[16]. However, lower residual stenosis in the SB has been observed with the two-stent strategies[27]. Comparing the two-stent methods, CRU and CUL were found to provide no significant difference in a three year follow-up study, with MACE outcomes occurring in 20.6% and 16.7% of patients, respectively.[19] The brief clinical results presented here demonstrate that clinical studies to date have been unable to provide conclusive evidence as to which stent technique performs best. For clinical trials, it is often impossible to differentiate adverse outcomes arising from the stent implantation techniques over all other clinical explanations. Additionally, since PSB is the preferred method, two-stent strategies are generally used only in more critical cases, thus biasing clinical results towards PSB.

In the present study, we sought to compare the hemodynamic conditions associated with in-vitro testing of different stent implantation techniques. This investigation expands upon the earlier work by Raben et al.[25] who reported the first in-vitro experimental results for the hemodynamics of stented coronary bifurcations using steady flow conditions. Here, we use particle image velocimetry (PIV) to obtain velocity and pressure fields under physiological pulsatile flow conditions for each coronary stent implantation technique.

## 2. Materials and Methods

### 2.1 Flow Loop

A mock circulatory flow loop was designed to simulate coronary flow conditions (refer to Charonko et al.[6] for more details). The working fluid, a 60/40 water to glycerin mixture, was used to match the kinematic viscosity and density of blood ($v = 3.77 \times 10^{-6}$ m$^2$/s, $\rho = 1100$ kg/m$^3$). Figure 1(a) illustrates the flow loop schematic. A pulsatile waveform, shown in Figure 1(b), was generated through the flow loop using a computer controlled gear pump. The mean flow rate was maintained around 85 mL/min for all test cases, modeling a resting flow condition with a heart rate of 70 bpm.[5] The flow rate was maintained at an 84/16 split between the MV and SB[25]. The pressure waveform, shown in Figure 1(b), was out of phase with the flow rate, mimicking the hemodynamic environment in the coronary artery.

Figure 1

## 2.2 Stent Models

Five compliant coronary artery models with a 60° bifurcation were cast using PDMS, as described by Raben et al.[25] The lumen diameters of the MV and SB were 3.96 mm and 2.77 mm, respectively. Commercially available Endeavor Resolute stents (Medtronic, Minneapolis, MN, USA) were implanted into three of the models by an interventional cardiologist, as they would be during a typical clinical procedure. Table 1 describes the implantation sequence used for each method.

Table 1

Figure 2 shows the stent models used in this experiment with a schematic of each implementation method tested.

Figure 2

## 2.3 PIV Setup

PIV images were captured using an Nd-YAG laser and a high-speed camera (IDT Xs-5i).[25] To match the index of refraction, the area surrounding the test section was also filled with the working fluid. The working fluid was seeded with 7 μm fluorescent particles. A frame pair frequency of 250 Hz was used with 200 μs between the images. The images were captured in the transversal plane of the vessel, of size 1728 x 2352 pixels, with a resolution of 7.04 µm/pixel for all stented cases and 7.73 µm/pixel for the un-stented case. Four cardiac cycles were acquired for each test case. PIV images were processed using an in-house PIV software Prana (https://github.com/aether-lab/prana). A multi-frame approach for dynamic range enhancement[14] and robust phase correlation algorithms for increased accuracy were used.[13,11,12] Four PIV passes were used, with the final pass using a 64 x 32 pixel window and 16 x 16 grid resolution. Proper orthogonal decomposition with a 90% energy cutoff was used on the time-resolved PIV velocity fields to reduce the effects of random errors.[18] To obtain a general quantification of the velocity uncertainty, the peak to peak ratio was analyzed, similar to the method described in Raben et al.[24], with further details in Charonko et al.[4] From this analysis, uncertainty in the PIV velocity fields is approximated at 4% and 13% for the un-stented and stented cases, respectively. The un-stented case had lower uncertainty since there were no stents to block the any particle motions in the image. A more rigorous uncertainty quantification for this experimental setup was done in Raben et al.[25], but is beyond the scope of this work.

## 2.4 Post-Processing

The PIV velocity fields were phase averaged. Masks of the stent locations were created using a connected components algorithm with 4-point neighborhoods and a 10-pixel threshold (refer to Sklansky[26] for more details on connected components algorithms). The location of the stent was computed in 25 image increments (1/20$^{th}$ of the pulsatile cycle) to account for the stretching and compressing of the stent through the pulsatile cycle. Velocity components that overlapped with stent struts were excluded from all post-processing calculations.

Particles near the vessel inlet in the PMV for all stented case were observed to be out of focus, hindering the velocity correlation in this region. Resultantly, velocities at the inlet of the PMV could not be resolved, as evident in the velocity fields shown in Figure 3. For this reason, we have removed this poorly resolved portion of the flow from all post-processing calculations.

Reduction of the centerline velocity was computed as the percent decrease in centerline velocity magnitude at peak flow rate for each stent case as compared to the un-stented case. The percent of vectors within 50% of the maximum velocity vector through time was also computed as a representative measure of the spatially varying momentum deficit induced by the stents. This metric also indicates an alteration of the velocity profile through the vessel.

Recirculating flow areas were identified by the angle of a velocity vector compared to a 0° and 60° reference angle in the MV and SB, respectively. Any vector that deviated by more than 20° from its reference angle was considered an indicator of recirculating flow. Degree of recirculation was defined as the number of time steps in which a vector was identified as maintaining recirculating flow, over the total number of time steps. Thus, one indicates the flow at that point is recirculating throughout the entire cycle while zero indicates flow in that region is never recirculating. Recirculation areas identified at the ostium of the SB were removed, since the velocity angle at that location should not abide by the 0° or 60° reference angle.

Time averaged WSS (TAWSS), Oscillatory Shear Index (OSI) and Relative Residence Time (RRT), given in Equations 2.1, 2.2, and 2.3 respectively, were computed in the MV.

$$TAWSS = \frac{1}{T} \int_0^T |\tau_w| dt \qquad (2.1)$$

$$OSI = \frac{1}{2}\left(1 - \frac{\frac{1}{T}\left|\int_0^T \tau_w dt\right|}{\frac{1}{T}\int_0^T |\tau_w| dt}\right) \qquad (2.2)$$

$$RRT = \frac{1}{(1-2OSI)TAWSS} \qquad (2.3)$$

where $\tau_w$ is the WSS vector and T is the duration of the cardiac cycle. To compute TAWSS, walls of the MV and SB in the test section were linearly defined and velocity gradients were obtained using thin-plate spline radial-basis functions (TPS-RBF) to decrease errors in the calculation.[18] It should be noted that the velocity fields are two-dimensional and thus the TAWSS computed here is one-dimensional. Additionally, a temporal moving average using four data points was used to smooth the trends and minimize noise caused by the numerical differentiation. The TAWSS code was validated using synthetic Poiseuille flow images. OSI values range from 0 to 0.5 with 0 indicating a flow with no oscillatory flow and 0.5 indicating a purely oscillatory flow. Time and space averaged WSS, OSI, and RRT values were obtained by numerically averaging the spatially varying results from Equations 2.1, 2.2, and 2.3.

PIV pressure fields for each stented case were evaluated using an in-house Navier-Stokes pressure solver described in Charonko et al.[2] The velocity fields following proper orthogonal decomposition (70% energy), prior to phase averaging, were used to compute the pressure in order to minimize errors. A pressure transducer just upstream of the geometry was used as the reference pressure for the code.

Subsequently pressure wave speeds 'c', as a representative measure of the stent compliance, were computed in the distal MV (DMV) and the SB using the following equation:

$$c = \frac{1}{\rho}\sqrt{\frac{\sum dP^2}{\sum dU^2}} \qquad (2.4)$$

where $\rho$ is the fluid density, and P and U are the instantaneous pressure and velocity, respectively. Using instantaneous pressure and velocity measurements at a single point reduces the effect of wave reflections on the calculation.[9] Additionally, it was shown that stent design does not have an effect on pressure wave reflections and thus the reflection magnitude should be similar for all cases.[3] An increase in pressure wave speed following stent implantation indicates a decrease in compliance and thus a larger compliance mismatch. Compliance mismatch is known to increase RRT, adversely alter the WSS distribution, and increase the risk of stent failure.[28]

## 3. Results

Table 2 provides a concise summary of all results presented here. This includes, reduction of centerline velocity, representative momentum deficit, TAWSS, OSI, RRT, and pressure wave speed. Risk factors for each hemodynamic parameter were computed for all stent implantation methods. Risk factors are considered to be the adverse percent change of a hemodynamic parameter caused by the stent implantation method as compared to the un-stented case. OSI, RRT, and pressure wave speed are the hemodynamic parameters that differentiate the stent implantation methods the most.

Table 2

The velocity magnitude fields at peak flow-rate (~200 mL/min) for each test case are shown in Figure 3. An immediately observable consequence of stent implantation is the reduction of centerline flow velocity in the DMV. The peak flow rate centerline velocity magnitude for the un-stented case was 0.68 m/s (Figure 4(a)). CRU provided the smallest centerline velocity reduction of 7.2%. PSB and CUL yielded similar velocity deficits of 15.7% and 18.4%, respectively. Further, CUL, and to a lesser extent PSB, demonstrate velocity profiles similar to that observed in the un-stented case. CRU altered the velocity profile in the DMV, skewing the centerline velocity towards the non-bifurcating wall.

Figure 3

Figure 4

Figure 4 further details the velocity changes in the MV for each stented case as compared to the un-stented case. In Figure 4(a), the maximum velocity magnitude in each time field through one cycle is plotted. From this figure, it is evident that each stenting method changes the time in the cycle when peak flow rate occurs. For the un-stented case the cycle peak occurs at time 0.274 while it occurs at time 0.256, 0.280, and 0.328 for PSB, CUL, and CRU, respectively. Figure 4(b) plots the percent of the MV velocity vectors at each time step that are within 50% of the maximum velocity magnitude at the given time step. The un-stented case averages 55.3% of vectors within 50% of the maximum velocity, while PSB, CUL, and CRU maintain averages of 25.4%, 29.4%, and 25.3% of vectors, respectively. This indicates that all stenting methods induce a large momentum deficit in the MV. Figure 4(c) shows the velocity fields at peak flow rate, normalized by the respective maximum velocity magnitude for each stent and masked to only show vectors within 50% of the maximum. CRU demonstrates a localized jet-like flow in the DMV, suggesting low flow exists near the walls. PSB and CUL exhibit wide velocity fields at peak flow rate and thus favorably low velocity profile narrowing in the DMV. PSB maintains an asymmetrical velocity profile, skewed towards the bifurcating wall, in the DMV.

Figure 5

Figure 5(a) shows the reduction of centerline velocity in the SB. The maximum velocity in the SB in the un-stented case is 0.50 m/s. PSB, CUL, and CRU induce a reduction of the maximum SB velocity by 43.9%, 58.4%, and 50.2%, respectively. While all stent cases yield large deficits of velocity magnitude, they produce a broader jet of flow into the SB. Figure 5(b) illustrates this with the percent of SB velocity vectors within 50% of the maximum SB velocity through time. The un-stented case maintains the lowest percentage of vectors with an average of 13.5%. CRU sustains a similar average of vectors of 13.9%. PSB and CUL, however, provide increases with 24.7% and 18.3% percent of vectors within 50% of the maximum velocity, respectively. Figure 5(c) shows the SB peak flow rate velocity fields, normalized by the respective maximum velocity for each stent and masked to only show vectors within 50% of the maximum. This further exhibits that all stent methods widen the jet of flow into the base of the SB as compared to the un-stented case. This also suggests that the stents partially mitigate the adverse hemodynamic effects of low velocity and recirculation at the proximal side of the SB caused by the high bifurcation angle.

Figure 6

To confirm the observations that the stented models attenuate the recirculating regions in the SB, Figure 6 shows the recirculation areas for each test case. Vectors with recirculating flow for less than 25% of the time were masked out in order to better visualize the regions of interest. Recirculating flow generally can cause low flow velocity, increased OSI and RRT, and higher risk of restenosis. All stent cases eliminate the large recirculation zone observed in the proximal side of the SB base of the un-stented case. The fact that this change, as well as the widening of the SB in-flow jet, is consistent across all implantation types suggests that this positive result may be due to the enlarged ostium of the bifurcation induced by the FKB procedure. The un-stented case and CRU show low velocity recirculating flow immediately following the SB on the bifurcating DMV wall. CUL shows a smaller and weaker recirculation region in this area. The small recirculating flow regions highlighted on the walls of all test cases are the result of low flow velocity near the walls combined with the unsteady nature of the pulsatility.

While recirculation zones are generally unfavorable, TAWSS, OSI, and RRT must be examined to determine the adverse risk that each zone causes.

Figure 7

Figure 7 shows the TAWSS for each case along the MV bifurcating and non-bifurcating wall. In the PMV, the TAWSS is notably low for all stented cases.

Previous studies have indicated that low TAWSS in the proximal MV can be the result of over-expansion of the stent, requiring recovery of WSS in order to restore physiologic flow conditions.[6,25,22] On the non-bifurcating wall, PSB and the un-stented case show a decreasing TAWSS trend in the DMV, a direct result of the asymmetric velocity profile in this location. Immediately following the SB on the bifurcating MV wall, the un-stented case and CRU both exhibit low TAWSS, a result of the recirculation zones observed in this location. Time and space averaged WSS values are given in Figure 8. CUL provided the smallest reduction of time and space averaged WSS of 17.1% as compared to the un-stented case. PSB and CRU yielded reductions of 31.4% and 35.3%. This reduction of overall time and space averaged WSS is due to the hemodynamics in the PMV where all stented methods yielded deficits of over 50%. In the DMV, CUL actually increased the time and space average WSS as compared to the un-stented case by 28.5%, while PSB and CRU maintained mild reductions of 5.5% and 13.6%, respectively.

Figure 8

OSI and RRT distributions did not show significant space-dependent trends through the MV and thus are not shown here. Time and space averaged values of OSI and RRT are given in Figure 8. In the DMV, CUL and CRU reduce the OSI by approximately 31% and 21%, respectively, as compared to the un-stented case. Meanwhile, PSB increases average OSI in the DMV by 33%, suggesting a detrimental effect of the high bifurcation angle persists with PSB in the DMV. In the PMV, PSB and CRU increase OSI by 473.3% and 115.5%, respectively. CUL maintains a significantly lower OSI increase in the PMV of only 47.3%. All stent cases increase the RRT of the vessel. Particularly, in the PMV as compared to the un-stented case, CUL provides a 2-fold increase in RRT, while PSB yields a 5-fold increase in RRT values. This is likely a consequence of the stent over-expansion and low TAWSS at this location. In the DMV, the stented cases maintain similar RRT results to the un-stented case.

Table 3

Table 3 reports the pressure wave speeds in the DMV and SB for each case. As expected, the implantation of the stent stiffens the vessel, thus increasing the pressure wave speed. CRU best approximates the un-stented vessel compliance for both the DMV and SB. CRU increases the pressure wave speed in the DMV by 55.5% while PSB and CUL yield increases of 164.8% and 113.0%, respectively. Thus, in the DMV, CUL provides increased performance as compared to PSB. In the SB, CUL and PSB are within the uncertainty bounds of the calculation and thus are considered equivalent. PSB increases pressure wave speed in the SB by 271.8% despite not having a stent implanted in the SB, suggesting that the FKB technique may adversely contribute to a compliance mismatch following stent implantation.

## 4. Discussion

Coronary branching with bifurcation angles over $50°$ are recognized to have higher risk of stenosis as they induce detrimental hemodynamic patterns[10]. In the un-stented case, the high bifurcation angle causes a large recirculation zone in the proximal side at the base of the SB. Additionally, the SB causes a centripetal acceleration of the flow pulling it upward and creating a slightly asymmetric velocity profile in the DMV. This causes low velocity flow on the non-bifurcating wall in the DMV and induces low and decreasing TAWSS at this location. Additionally a low flow region is present immediately following the SB on the bifurcating wall in the DMV, as observed in Figure 6. This causes low TAWSS at the start of the DMV in this location. These hemodynamic observations cultivate three high-risk zones that are susceptible to stenosis: (1) the large recirculating region in the SB, (2) the low flow region near the carina, and (3) the non-bifurcating wall in the DMV. A successful stenting procedure aims to restore normal hemodynamic conditions through a vessel by reopening an occluded vessel. However, cases exhibiting a high bifurcation angle present a unique challenge because "normal hemodynamic conditions", even without a stent, maintain adverse hemodynamic conditions. For this reason, in cases such as the one presented here, where a high bifurcation angle exists, the stent implantation procedure seeks to restore blood flow to normal hemodynamic conditions while also mitigating the natural and deleterious effects of the high bifurcation angle. From the results presented here, it is evident that each stent implantation method achieves these two goals with varying success, as each technique produces different hemodynamic environments.

All stenting methods are able to eliminate the large recirculation zone observed in the proximal side of the base of the SB in the un-stented case. As previously mentioned, the FKB procedure widens the ostium of the bifurcation, yielding a more gradual transition from the PMV to the SB. This gradual transition is observable by examining the geometry and stent outlines in Figure 6. For both PSB and CUL, because the recirculation region in the SB is eliminated with all stenting methods, the effective area of flow into the SB is increased, in accordance with Figure 5(b). Thus, to maintain continuity, the velocity magnitude at the ostium of the SB must decrease. This elucidates the reduction of maximum velocity into the SB at peak flow rate by 50% or more in all stented cases, as observed in Figure 5(a). With CRU, despite the reduction of maximum velocity into the SB at peak flow rate, a high momentum deficit persists in the SB, since CRU and the un-stented case maintain a similar percentage of vectors in the SB within 50% of the maximum. This is because CRU has a high strut density in the PMV on the bifurcating wall just before the SB.

Because of the low flow velocity and unsteadiness in the pulsatile waveform, small eddies are produced near the wall when the bulk flow velocity is low as evident by the recirculation regions indicated along the walls of all test cases in Figure 6. This is also the explanation for the recirculation region indicated at the carina region with the un-stented case. However, in the case of CRU, the recirculation zone is an artifact of high strut density and interaction at that location, resulting in flow disturbances. Thus, this demonstrates an adverse hemodynamic outcome of the stent configuration with the CRU method as it causes a significant flow disturbance near the bifurcation, both before and after the SB. Subsequently, with CRU, the TAWSS drops following the SB and must increase throughout the length of the DMV non-bifurcating wall in order to restore the flow conditions following the strut induced flow disturbances. CUL also shows a low flow region near the carina, though considerably smaller than with CRU. This is because CUL maintains overlapped struts at this location, but the two stents have a more limited interaction than with CRU. Additionally, CUL does not cause low TAWSS to persist for any length on the bifurcating wall in the DMV as evident in Figure 7, indicating that the low flow area maintains minimal hemodynamic disturbances. PSB does not show a low velocity flow region at the carina level, in accordance with previous CFD results.[7,22,25]

PSB was previously noted to maintain an asymmetric velocity profile in the DMV. CUL and the un-stented case both maintain slightly skewed profiles in the DMV, but to a lesser extent than PSB. The DMV velocity profile produced by CUL is very similar to that of the un-stented case. This indicates that CUL best maintains a normal velocity profile in the DMV. Because PSB does not utilize a stent in the SB, it is unable to widen the ostium in the same manner the as the two stent methods. Thus, PSB maintains a stronger centripetal force than CUL or CRU, yielding a larger upward force on flow in the DMV and a velocity profile more skewed towards the bifurcating wall. This results in low flow velocity on the non-bifurcating wall, explaining why PSB yields higher OSI and RRT in the DMV than both CUL and CRU.

To the authors' knowledge, this is the first study to examine compliance mismatch to compare coronary bifurcation stent implantation techniques. Further, examination of the pressure wave speeds demonstrates that the stenting methods can actually induce different levels of compliance. CRU provided the smallest compliance mismatch throughout the vessel, revealing a major advantage for the CRU technique. Adversely, PSB provided the largest and most adverse decrease in compliance of the MV. The large compliance mismatch may also contribute to the increase of OSI and RRT values seen with PSB in Figure 8. The high compliance mismatch and OSI and RRT values represent major disturbances of normal flow conditions that PSB induces.

Overall, this study demonstrated both positive and negative hemodynamic effects observed with all implantation methods. CRU provided some advantages, most notably the lowest compliance mismatch. However, CRU demonstrated the lowest TAWSS average and an adverse jet-like velocity profile in the DMV. Additionally, with CRU, we recognized a disadvantage associated with the interaction of the two stents resulting in high flow disturbances in the MV near the carina. Despite its simplicity, PSB showed several favorable hemodynamic results including the elimination of major recirculation zones and widening of the SB inflow jet. However, PSB yielded the highest and most adverse OSI and RRT averages and MV compliance mismatch. Meanwhile, CUL provided a balanced hemodynamic environment that eliminated the adverse effects of the high bifurcation angle and showed many indications of maintaining normal flow conditions. It yielded time and space averaged WSS, OSI, and RRT values that most closely matched that of the un-stented case. Overall, CUL provided the most synergistic stenting solution. Despite utilizing two stents, CUL yields minimal stent induced flow disturbances. Additionally, disruptions of the flow that are observed with CUL do not propagate into TAWSS, OSI, or RRT. Thus, these results demonstrate that both PSB and CUL are able to retain many aspects of normal flow conditions with minimal flow disturbances. However, CUL mitigated the detrimental effects induced by a high bifurcation angle, while PSB fell short.

This study, as with all experimental studies, has limitations. WSS, OSI, and RRT calculations on the walls of the SB were subject to experimental noise and stent interference. Additionally, the results presented here are constrained to one plane of the bifurcation, making overall distributions of TAWSS, OSI, and RRT unknown. The results also do not account for factors such as overlapping stent struts that increase risk of mechanical stent failure.[23] Therefore, while the results presented here indicate the major hemodynamic differences between the stent implantation methods, final conclusions and comparisons between the stent cases must be taken with caution, as the experimental limitations impose an inability to directly predict clinical outcomes.

## 5. Conflict of Interest statement

## 6. Acknowledgements

The authors would like to thank Jaime S. Raben for her contributions to the initial efforts of this project. Pavlos Vlachos acknowledges partial support by NIH NHLBI Grant No HL106276-01A1. Claudio Chiastra is partially supported by the ERC starting grant (310457, BioCCora).

## 8. Figures and Tables

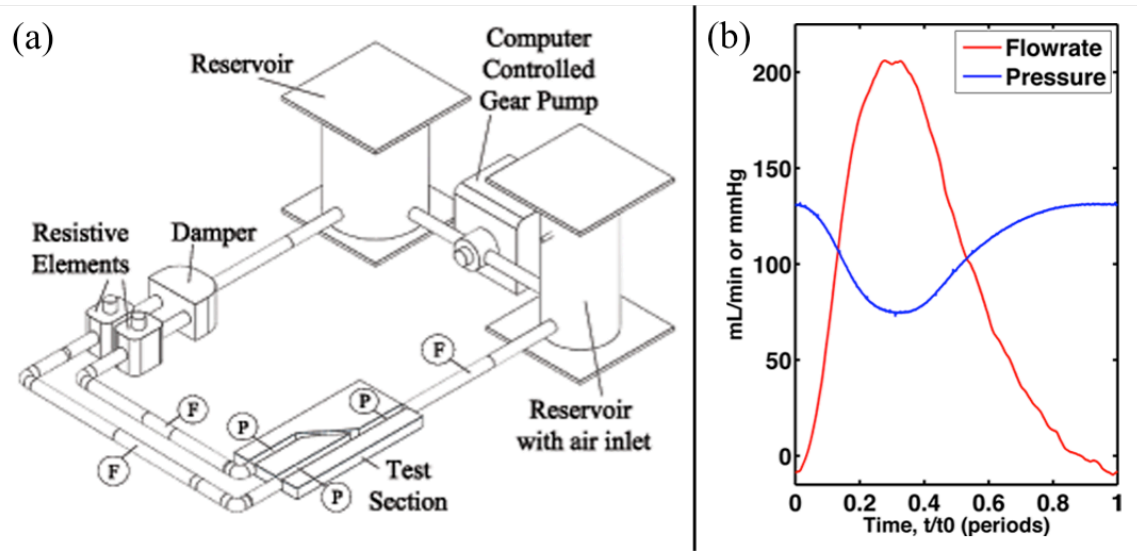

Figure 1: (a) Mock coronary flow loop schematic (adapted from Raben et al.[25]), (b) Flow rate and pressure through loop

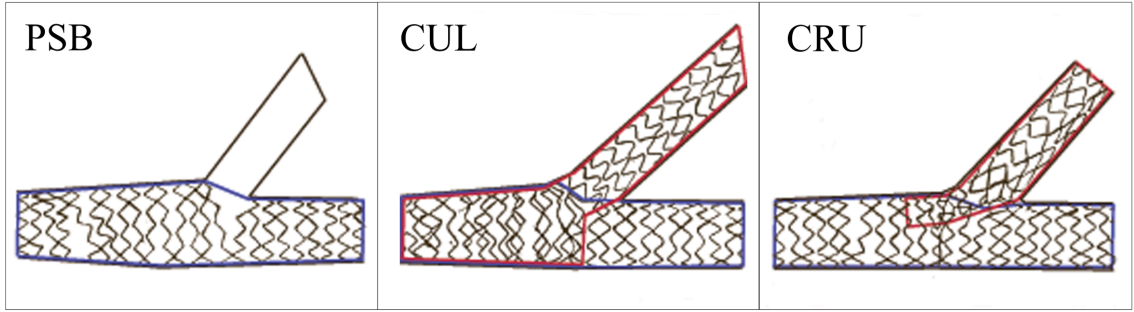

Figure 2: Provisional side branch (PSB), culotte (CUL), and crush (CRU) stented models. MV stents are outlined with blue, SB stents are outlined in red. (Adapted from Raben et al.[25])

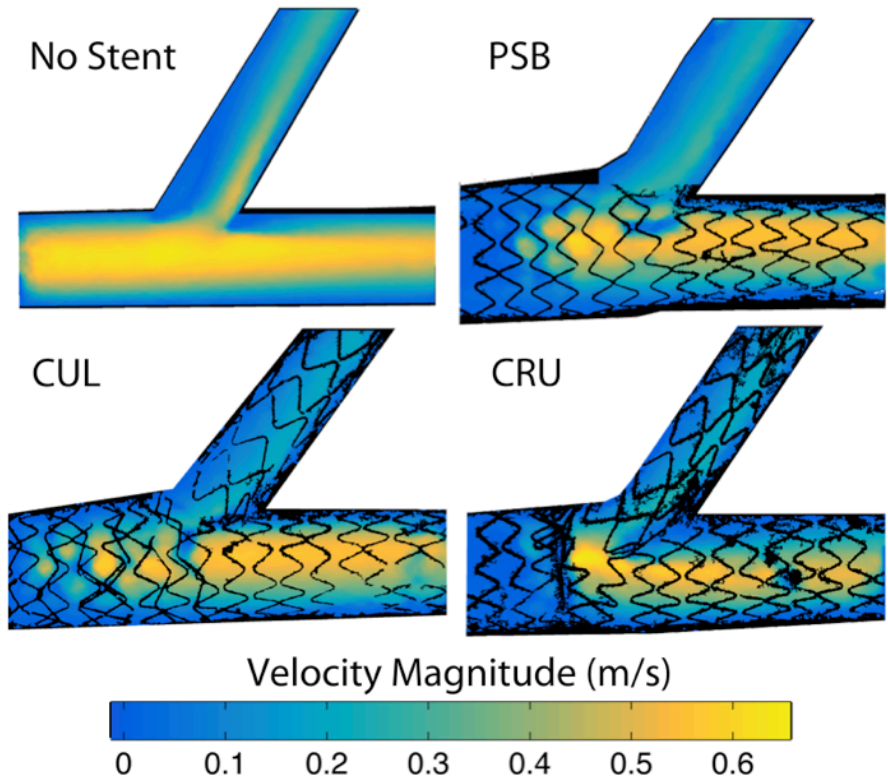

**Figure 3: Velocity magnitude of each test case at peak flow rate**

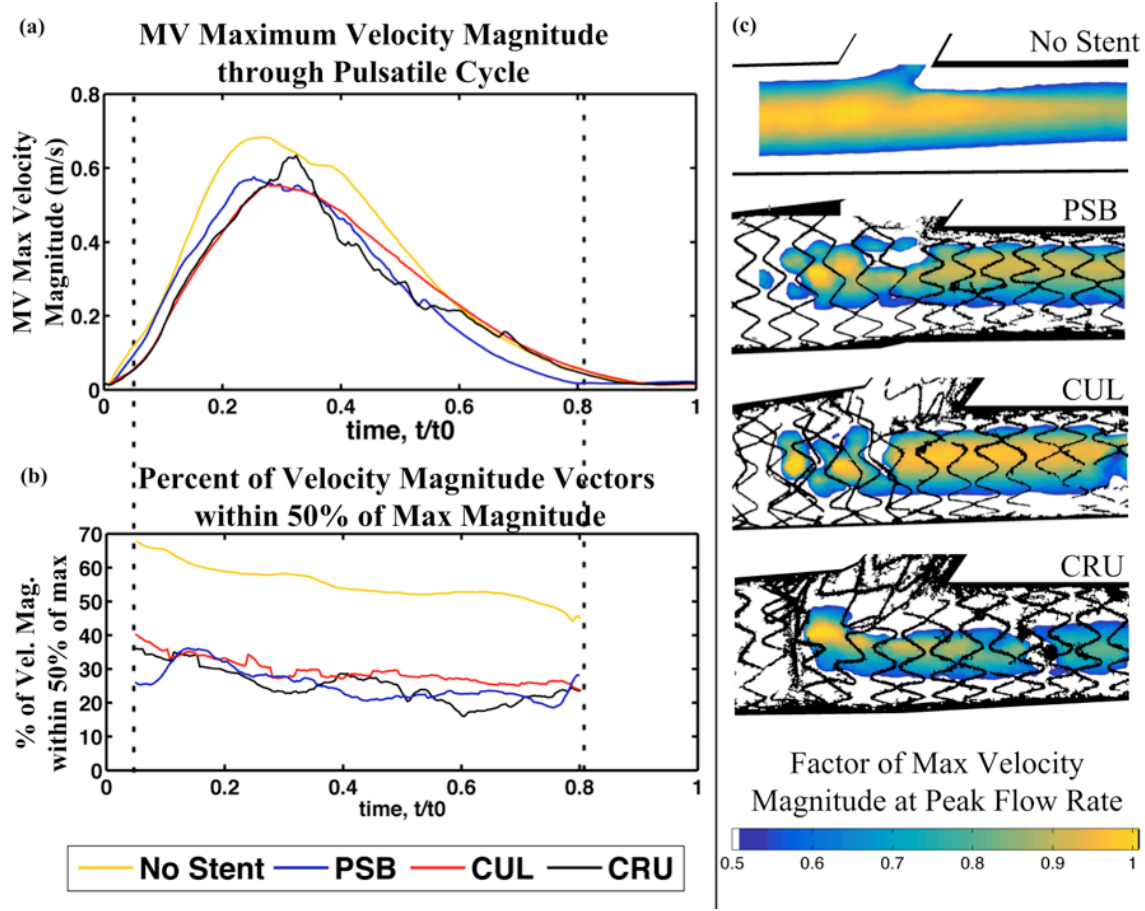

Figure 4: (a) Maximum velocity magnitude in MV for each test case through pulsatile cycle (b) Velocity vectors at a given time in MV that are within 50% of maximum velocity at that time, (c) Normalized velocity vectors in MV at peak flow rate that are within 50% of the maximum velocity vector.

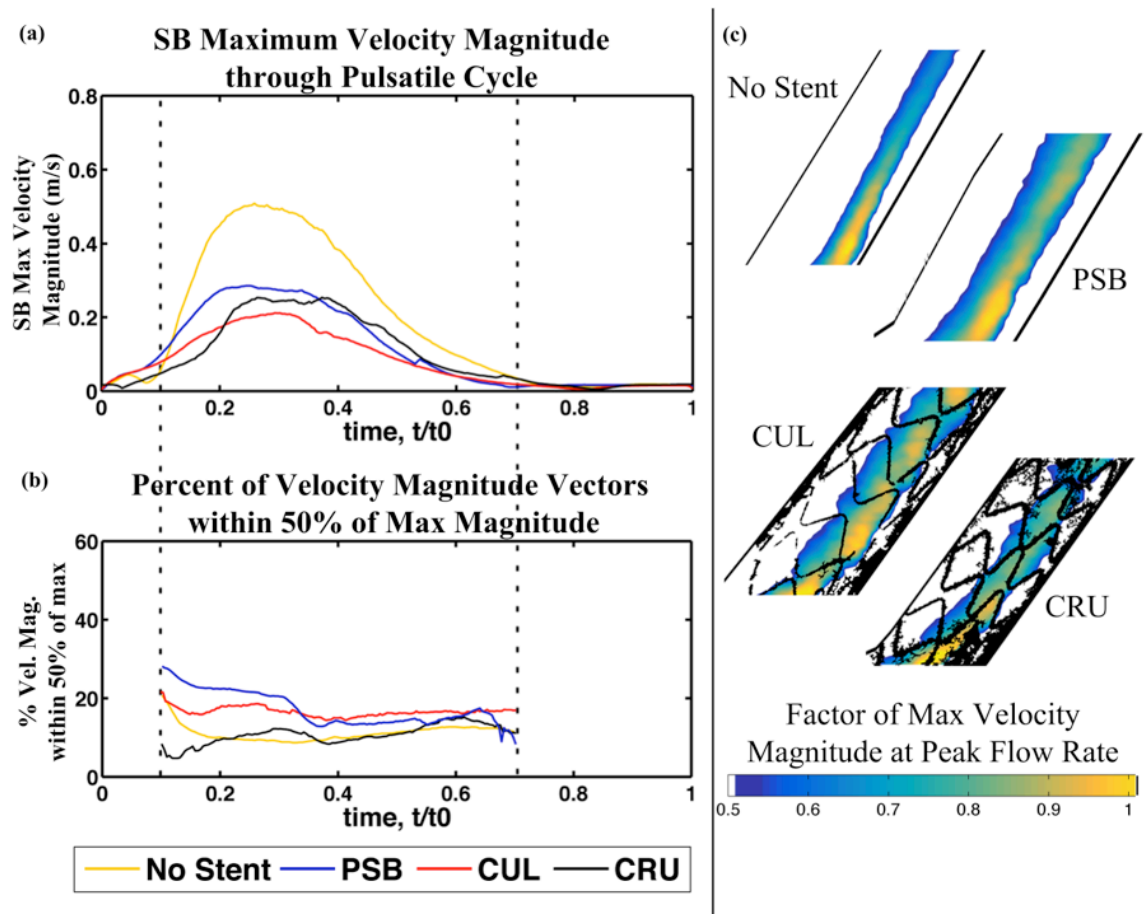

Figure 5: (a) Maximum velocity magnitude in SB for each test case through pulsatile cycle (b) Velocity vectors at a given time in SB that are within 50% of maximum velocity at that time, (c) Normalized velocity vectors in SB at peak flow rate that are within 50% of the maximum velocity vector.

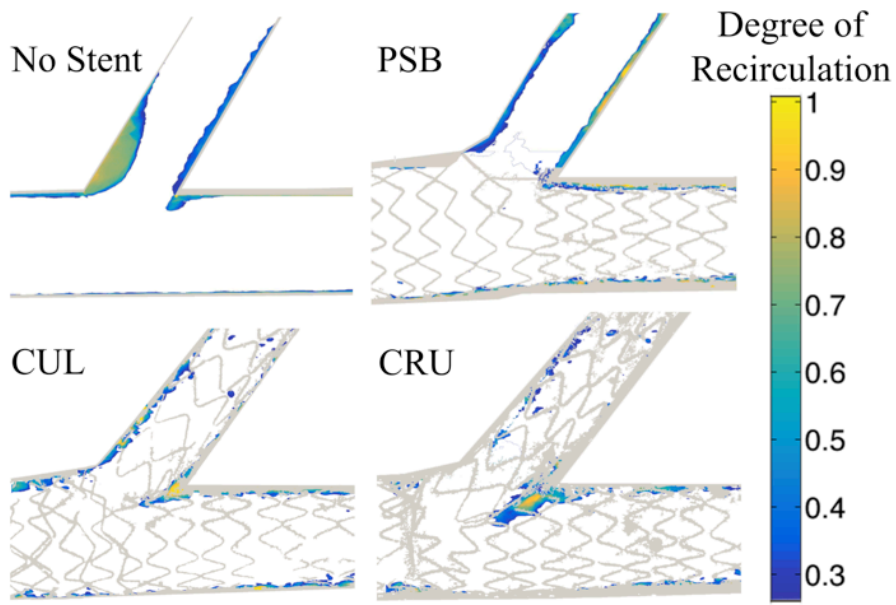

Figure 6: Recirculation regions for each test case, where degree of recirculation represents the percentage of time within the pulsatile cycle that the flow deviates by 20° or more from a reference angle (0° in the MV or 60° in the SB). Stent mask is included to show relative locations.

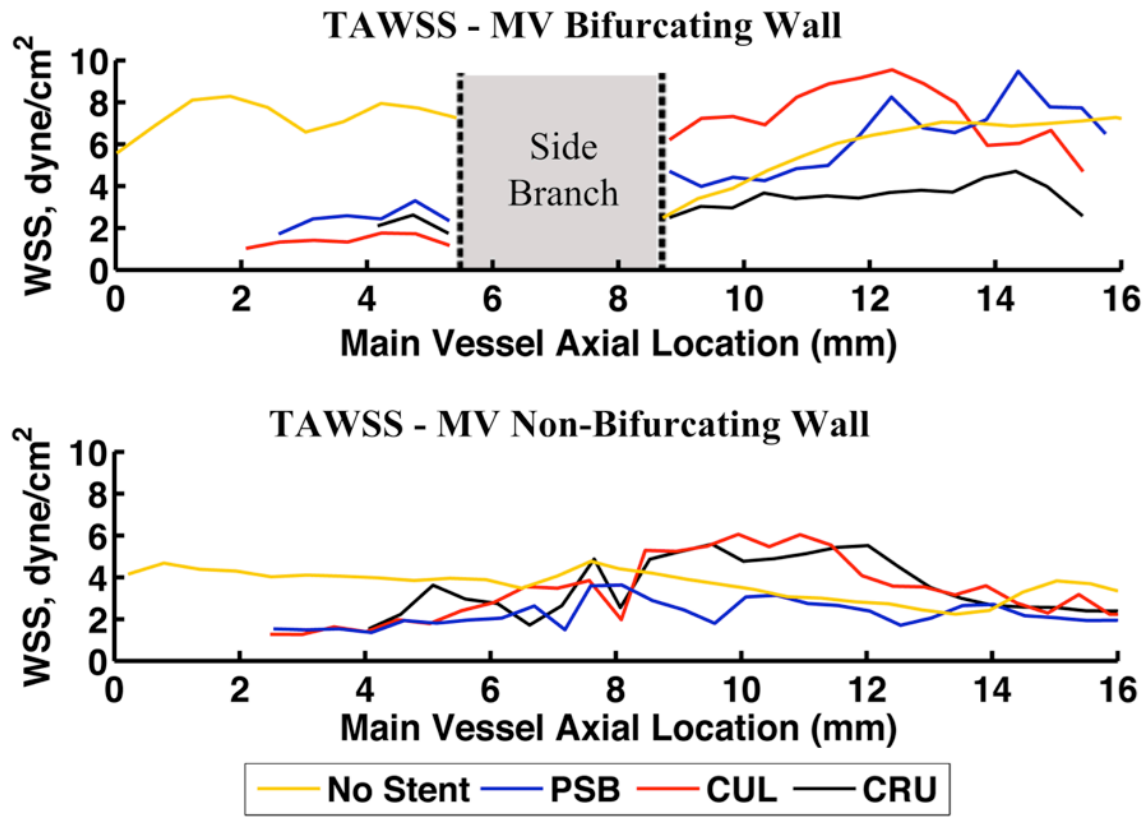

**Figure 7:** Time averaged wall shear stress in the MV for each test case using a four-point moving average to smooth noise from differentiation.

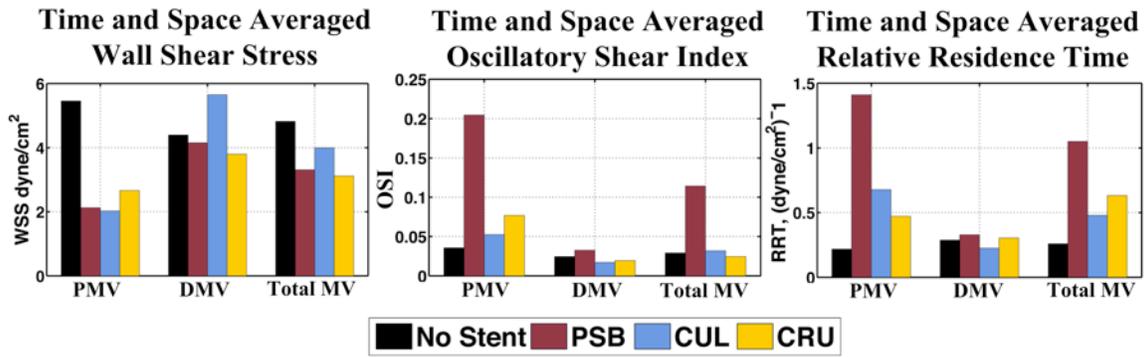

Figure 8: Time and space averaged wall shear stress, oscillatory shear index, and relative residence time in the Proximal MV, Distal MV, and MV.

**Table 1: Step by step sequence of each stent implantation method used (FKB = Final Kissing Balloon, POT = Proximal Optimization Technique)**

| PSB | CUL | CRU |
|---|---|---|
| 1. MV stent implantation | 1. Stent implantation from SB to PMV | 1. SB stenting with small protrusion in MV |
| 2. POT (MV post-dilation) | 2. POT (MV post-dilation) | 2. SB stenting crush by MV balloon inflation |
| 3. FKB | 3. MV rewiring | 3. MV stent implantation |
| 4. POT (MV post-dilation) | 4. MV Dilation | 4. SB rewiring |
| | 5. MV stent implantation (distal to proximal) | 5. FKB |
| | 6. SB rewiring | 6. POT (MV post-dilation) |
| | 7. FKB | |
| | 8. POT (MV post-dilation) | |

Table 2: Results summary containing velocity, flow parameters, wave speeds, and risk factors for each stent implantation method. Risk factors were computed as the percent risk increase for each parameter

| Flow Property | | No Stent | PSB | CUL | CRU |
|---|---|---|---|---|---|
| **Velocity Data** | MV max velocity magnitude (m/s) | 0.683 | 0.576 | 0.557 | 0.634 |
| | SB max velocity magnitude (m/s) | 0.509 | 0.285 | 0.212 | 0.253 |
| | TA MV velocity vectors within 50% of max (%) | 55.3% | 25.4% | 29.4% | 25.3% |
| | TA SB vel. vectors within 50% of max (%) | 13.5% | 24.7% | 18.3% | 13.9% |
| **Time and Space Averaged Flow Parameters** | PMV - WSS (dyne/cm$^2$) | 5.46 | 2.13 | 2.02 | 2.66 |
| | PMV - OSI | 0.036 | 0.205 | 0.053 | 0.077 |
| | PMV - RRT (dyne/cm$^2$)$^{-1}$ | 0.22 | 1.41 | 0.68 | 0.47 |
| | DMV - WSS (dyne/cm$^2$) | 4.40 | 4.16 | 5.65 | 3.80 |
| | DMV - OSI | 0.025 | 0.033 | 0.017 | 0.019 |
| | DMV - RRT (dyne/cm$^2$)$^{-1}$ | 0.288 | 0.329 | 0.226 | 0.304 |
| | Total - WSS (dyne/cm$^2$) | 4.82 | 3.31 | 4.00 | 3.12 |
| | Total - OSI | 0.029 | 0.114 | 0.032 | 0.025 |
| | Total - RRT (dyne/cm$^2$)$^{-1}$ | 0.26 | 1.05 | 0.48 | 0.63 |
| **Pressure Wave Speeds** | DMV (m/s) | 5.4 | 14.3 | 11.5 | 8.4 |
| | SB (m/s) | 7.1 | 26.4 | 26.8 | 17.9 |
| **Stent Induced Hemodynamic Risk Factors** | MV centerline vel. | - | 15.8% | 18.5% | 7.2% |
| | SB centerline vel. | - | 43.9% | 58.4% | 50.2% |
| | WSS deficit (%) | - | 31.4% | 17.1% | 35.3% |
| | OSI increase (%) | - | 293.8% | 10.2% | -15.3% |
| | RRT increase (%) | - | 305.0% | 84.7% | 143.8% |
| | Pressure wave speed (DMV) increase (%) | - | 164.8% | 113.0% | 55.5% |
| | Pressure wave speed (SB) increase (%) | - | 271.8% | 277.5% | 152.1% |

**Table 3: Pressure Wave Speeds in the DMV and SB including uncertainty**

|  |  | No Stent | PSB | CUL | CRU |
|---|---|---|---|---|---|
| **Pressure Wave Speeds** | DMV (m/s) | 5.4 ± 0.3 | 14.3 ± 0.6 | 11.5 ± 0.6 | 8.4 ± 0.3 |
|  | SB (m/s) | 7.1 ± 0.5 | 26.4 ± 1.2 | 26.8 ± 1.4 | 17.9 ± 0.9 |